# Improved Upper Critical Field in Bulk-Form Magnesium Diboride by Mechanical Alloying With Carbon

*B. J. Senkowicz, J. E. Giencke, S. Patnaik\*, C. B. Eom, E. E. Hellstrom, and D. C. Larbalestier*

*Applied Superconductivity Center and Department of Materials Science and Engineering, University of Wisconsin – Madison, Madison, WI 53706*

\*On leave from School of Physical Sciences, J. Nehru University, New Delhi 110067, India

High-energy milling of magnesium diboride ($MgB_2$) pre-reacted powder renders the material largely amorphous through extreme mechanical deformation and is suitable for mechanically alloying $MgB_2$ with dopants including carbon. Bulk samples of milled carbon and $MgB_2$ powders subjected to hot isostatic pressing and Mg vapor annealing have achieved upper critical fields in excess of 32T and critical current density approaching $10^6$ A/cm$^2$.

Recent studies of magnesium diboride thin films by Braccini et al.[1] found $H_{c2}(0K)^{//} > 50T$ for C-doped MgB$_2$ films. Fits to the data using the two-gap model of $H_{c2}$ of Gurevich[2] showed an extrapolated $H_{c2}(0K)^{//}$ (H parallel to the Mg and B planes) of ~ 70T and $H_{c2}(0K)^{\perp}$ exceeding 40T [1]. Such critical field properties exceed those of any Nb-base conductor at any temperature, suggesting that MgB$_2$ could be a viable replacement for Nb$_3$Sn as a high field magnet conductor. Masui et al[3], Ohmichi et al.[4], and Kazakov et al.[5] have reported carbon-doped single crystals with $H_{c2}(0K)^{//}$ in excess of 30T. Untextured carbon-doped filaments fabricated by a CVD (chemical vapor deposition) method can achieve upper critical fields in excess of 30T at 4.2K[6]. The measured $H_{c2}$ in this case is the higher parallel critical field $H_{c2}^{//}(0K)$, rather than the lower, perpendicular critical field which controls the bulk $J_C$. One immediate question is how thin films and bulk differ. A partial answer was given by Braccini et al.[1] who noted that the c-axis parameter of the highest $H_{c2}$ films was expanded relative to C-doped bulk samples, which have lower $H_{c2}$ values.

Several groups[7-10] have also produced wires containing SiC, the latter finding that $H_{c2}$ could attain values as high as the CVD filaments of Wilke et al.[6] In fact there is increasing suspicion that one of the beneficial effects of SiC addition occurs by C doping of the MgB$_2$. In general there is good agreement that C-doping of MgB$_2$ is a useful means of alloying MgB$_2$, even if the best bulk $H_{c2}$ values are only about half those of the best thin films. The present letter discusses the ex situ synthesis of alloyed MgB$_2$ powder using high energy ball milling of MgB$_2$ with C. Since a major goal of MgB$_2$ technology is the fabrication of high critical current density, multifilament wire suitable for magnet applications, we need a scalable bulk process capable of producing carbon-doped precursor powder. One such method is provided by this work.

The composition of carbon-doped MgB$_2$ is commonly given using the format Mg(B$_{1-X}$C$_X$)$_2$. Alfa-Aesar pre-reacted MgB$_2$ powder was mixed with powdered graphite in the following proportions: 0.1C + 0.9MgB$_2$ (nominal X=0.0525) and 0.3C + 0.7MgB$_2$ (nominal X=0.17). The mixtures were high energy ball milled in a Spex 8000M mixer/mill with WC milling container and milling media. Powders were milled until no graphite peak was discerned in the x-ray diffraction pattern, which took 10 hours for X=0.0525 and 25 hours for X=0.17. The milled powders were then cold isostatic pressed to form pellets that were welded into evacuated stainless steel tube and hot isostatic pressed (HIP) at 1000°C and >30ksi for 200 minutes. These HIP-treated pellets were cut apart and examined. They were black, lusterless, and lacked mechanical strength. These features were attributed to incomplete reaction due to a deficiency of Mg with respect to (B+C) resulting from adding C to the ex-situ powder without adding Mg. Accordingly, portions of the HIP treated pellets were welded into evacuated stainless steel tubing with a quantity of Mg metal that was about half the volume of the MgB$_2$ pellet. MgB$_2$ pellets and Mg metal were arranged in such a way that the sample would be exposed to Mg vapor, while limiting exposure to liquid Mg, as described by Braccini et al.[11] These sealed MgB$_2$ + Mg packages were heat treated at 900°C for five hours and cooled at a rate not faster than 100°C/hr so as to fully react the B and C with Mg. After cutting the samples free from the steel, the samples were found to be black in color, to gain luster when polished, and to be dense and hard. However, even after this Mg vapor anneal, the x-ray intensities were reduced and the peaks broadened as compared with unmilled powders.

Some superconducting properties were then evaluated resistively with small transport currents of ~0.25 A/cm$^2$ in a 9T Quantum Design physical properties measurement system (PPMS) and then with the 65 Tesla short pulse magnet at the National High Magnetic Field Lab

(NHMFL) in Los Alamos, NM. $H_{c2}(T)$ was defined by the 90% normal state resistance transition[11] giving values about 5-10% lower than would be obtained by the 95-99% transitions[6]. The hysteretic magnetization ΔM of bulk samples was also measured in an Oxford Instruments vibrating sample magnetometer (VSM), from which the critical current density $J_c(H,T)$ was calculated assuming fully connected samples using the expression $J_c(H,T) = \Delta M*12b/(3bd-d^2)$ where b and d are the width and thickness of a rectangular section bar.

Figure 1 compares $H_{c2}(T)$ of our samples to those of an X=0.035C filament of Wilke et al.[6]. Our X=0.0525 sample had $T_c$ of 35K and $H_{c2}(0K)$ ~33T, and our nominal X=0.17 sample had $T_c$ of 22K and ~19T $H_{c2}(0K)$. The CVD filament[6] exhibited very similar properties to our X=0.0525 sample, given that $H_{c2}(T)$ was determined from the highest fraction of the transition, rather than the 90% transition as here[6,12]. These data also agree well with single crystal data for which an estimated maximum $H_{c2}^{//}(0K)$ ~ 29T at X=0.05[3] was reported.

To determine how much carbon was successfully incorporated into the lattice, Figure 2 compares **a**-axis lattice parameters in the literature[5,13-15]. Both single crystal data sets (Kazakov et al.[5] and Lee et al.[15]) are rather linear in C content. Fitting the Kazakov et al.[5] data produces the following relationship:

$$a = 3.08439x - 0.3153 \quad (1)$$

where x is the carbon content given by $Mg(B_{1-X}C_X)_2$.

Lattice parameters determined from high resolution x-ray diffraction (XRD) measurements are presented in Table I along with the carbon content of the lattice calculated using equation (1). This calculation suggests that x=0.051 of the nominally X=0.0525 sample (Sample A) was incorporated into the lattice. For the nominal X=0.17 carbon-doped sample (Sample B), only about 40% of the available carbon was incorporated into the lattice (x = 0.069). Between X=0.0525 and X=0.17 we experienced diminishing efficiency of carbon incorporation, as previously observed by Bharathi *et al.*[14]

Figure 3 shows $J_C(H,T)$ of the samples. $J_C$ of the X=0.0525 sample had a maximum approaching $10^6$ A/cm$^2$, similar to values in the literature for HIP processed ex-situ MgB$_2$[16]. The X=0.17 sample had maximum $J_C$ depressed two orders of magnitude, with depressed irreversibility fields $H^*(T)$ as well (defined as the field at which $J_C$ vanishes). As noted by comparing Fig. 1 and 3, $H^*(T)$ is of order one half $H_{c2}(T)$, for example 6T at 20K versus 10T at 20K for the X=0.051 sample. This is consistent with $H_{c2}(T)$ being determined by the highest $H_{c2}(T)$ path in the samples which corresponds to $H_{c2}^{//}(T)$ whereas $H^*(T)$ is determined by the breakup of the fully connected path that starts to occur in the vicinity of $H_{c2}^{\perp}(T)$.

As shown in Table 1, the normal state resistivity values were two to three orders of magnitude higher than those measured in carbon-doped single crystals[5] or in good undoped samples[17,11]. The normal state resistivity of the X=0.17 sample ($\rho(40K) = 9000$ µohm-cm) was nearly a factor of ten higher than that of the X=0.0525 sample (960 µohm-cm). Such values are indicative of reduced connectivity[18], a conclusion supported by their essentially metallic temperature dependence and residual resistivity ratio (RRR) values (1.185 for X=0.0525 and 1.037 for X=0.17). It is interesting that the $J_c(H,T)$ values can still approach $10^6$ A/cm$^2$, even with this high resistivity. Like some SiC-doped wires of Dou *et al.*[8] with high $\rho(40K)$ values, it appears that the inferred poor connectivity of the samples is not a barrier to attaining high $J_c$ values. An additional intrinsic source of high resistivities in both samples is likely due to incomplete recovery of milling-induced disorder during heat treatment, as supported by the weak

XRD patterns with broad peaks. Such inhibited recovery is generally not observed in undoped samples, leading to the conclusion that the presence of carbon hinders eliminating defects after milling. The low $J_C$ of the X=0.17 sample and the conclusion that not all of the C is incorporated into the lattice indicates additional obstruction by residual C phases, as has been seen in some C-doped films [19].

In summary we have shown that milling C with $MgB_2$ can produce $H_{c2}(0K)$ equal to that obtained for single crystals and CVD filaments. Lattice disorder introduced in the milling process is indicated by weakened XRD patterns, high normal state resistivity, and a low-temperature upturn in $H_{c2}(T)$.

BJS was supported by the Fusion Energy Sciences Fellowship Program, administered by Oak Ridge Institute for Science and Education under a contract between the U.S. Department of Energy and the Oak Ridge Associated Universities. This research was also supported by the NSF under the University of Wisconsin- Madison MRSEC program. The authors thank the excellent staff of the NHMFL-Los Alamos, particularly Dr. F. Balakirev, as well as W. Starch, A. Squitieri, J. Mantei, and R. Mungall in Wisconsin.

1. V. Braccini, A. Gurevich, J. E. Giencke, M. C. Jewell, C. B. Eom, D. C. Larbalestier, A. Pogrebnyakov, Y. Cui, B. T. Liu, Y. F. Hu, J. M. Redwing, Qi Li, X. X. Xi, R. K. Singh, R. Gandikota, J. Kim, B. Wilkens, N. Newman, J. Rowell, B. Moeckly, V. Ferrando, C. Tarantini, D. Marre, M. Putti, C. Ferdeghini, R. Vaglio, and E. Haanappel, Phys. Rev. B 71, 012504 (2005)

2. A. Gurevich, Physical Review B, **67,** 184515 (2003).

3. T. Masui, S. Lee, A. Yamamoto, H. Uchiyama and S. Tajima, Physica C: Superconductivity and its Applications, **412-414,** 303-6 (2004).

4. E. Ohmichi, T. Masui, S. Lee, S. Tajima and T. Osada, J. Phys. Soc. Jpn. 73, 2065 (2004).

5. S. M. Kazakov, R. Puzniak, K. Rogacki, A. V. Mironov, N. D. Zhigadlo, J. Jun, Ch. Soltmann, B. Batlogg, J. Karpinski, Phys. Rev. B 71, 0245331 (2005).

6. R.H.T. Wilke, S.L. Bud'ko, P.C. Canfield, D.K. Finnemore, R.J. Suplinskas and S.T. Hannahs, Phys. Rev. Lett., **92,** 217003-1 (2004).

7. S. X. Dou, S. Soltanian, J. Horvat, X. L. Wang, S. H. Zhou, M. Ionescu, H. K. Liu, P. Munroe, M. Tomsic. Appl. Phys. Lett., **81,** 3419 (2002).

8. S. X. Dou, V. Braccini, S. Soltanian, R. Klie, Y. Zhu, S. Li, X. L. Wang, D. Larbalestier, J. Appl. Phys., 96, 12, 7549 (2004)

9. A. Matsumoto, H. Kumakura, H. Kitaguchi and H. Hatakeyama, Supercond Sci Technol, **17,** 319-23 (2004).

10. M. D. Sumption, M. Bhatia, M. Rindfleisch, M. Tomsic, S. Soltanian, S. X. Dou, and E. W. Collings, Appl. Phys. Lett. 86, 092507 (2005)

11. V. Braccini, L. D. Cooley, S. Patnaik, D. C. Larbalestier, P. Manfrinetti, A. Palenzona, A. S. Siri.  Appl. Phys. Lett., **81,** 4577-9 (2002).

12. R. Puzniak, M. Angst, A. Szewczyk, J. Jun, S.M. Kazakov and J. Karpinski, Cond. Mat.0404579 , (2004).

13. T. Takenobu, T. Ito, D.H. Chi, K. Prassides and Y. Iwasa, Physical Review B (Condensed Matter and Materials Physics), **64,** 134513-1 (2001).

14. A. Bharathi, S. J. Balaselvi, S. Kalavathi, G. L. N. Reddy, V. S. Sastry, Y. Hariharan, T. S. Radhakrishnan. Physica C, **370,** 211-18 (2002).

15. S. Lee, T. Masui, A. Yamamoto, H. Uchiyama and S. Tajima.  Physica C: Superconductivity and its Applications, **397,** 7-13 (2003).

16. A. Serquis, L. Civale, D. L. Hammon, X. Z. Liao, J. Y. Coulter, Y. T. Zhu, M. Jaime, D. E. Peterson, F. M. Mueller, V. F. Nesterenko, Y. Gu.  Appl. Phys. Lett., 2003, **82**, 17, 2847-9.

17. P. C. Canfield, D. K. Finnemore, S. L. Bud'ko, J. E. Ostenson, G. Lapertot, C. E. Cunningham, C. Petrovic.  Phys. Rev. Lett., **86,** 2423-6 (2001).

18. J.M. Rowell, Supercond. Sci. Technol. **16,** 17-27 (2003).

19. J. M. Redwing, A. Pogrebnyakov, S. Raghavan, J. E. Jones, X. X. Xi, S. Y. Xu, Qi Li; Z. K. Liu, V. Vaithyanathan, D. G. Schlom.  Appl. Phys. Lett., **85,** 2017-19 (13).

Figure 1- $H_{c2}$ defined by the 90% normal state resistivity of our carbon-doped samples and final onsets of full normal-state resistance for the sample of Wilke et al..[6]

Figure 2- a-axis lattice parameter as a function of carbon content. Data from Kazakov et al.[5], Takenobu et al.[13], Bharathi et al.[14], and Lee et al..[15]

Figure 3 – Magnetization $J_c(H,T)$ values derived for X=0.0525 (black lines) and X=0.17 (gray lines). Comparison of the fields at which $J_c$ tends to zero are about half the values of $H_{c2}(T)$ found in Fig. 1.

Table I – Properties of C-doped milled samples

| Nominal X | a (Å) | c (Å) | Calculated X | $\rho$ (40K) ($\mu\Omega$-cm) | RRR |
|---|---|---|---|---|---|
| 0.0525 | 3.0683(2) | 3.5267(5) | 0.051 | 960 | 1.185 |
| 0.17 | 3.062(3) | 3.526(7) | 0.069 | 9000 | 1.037 |

Figure 1

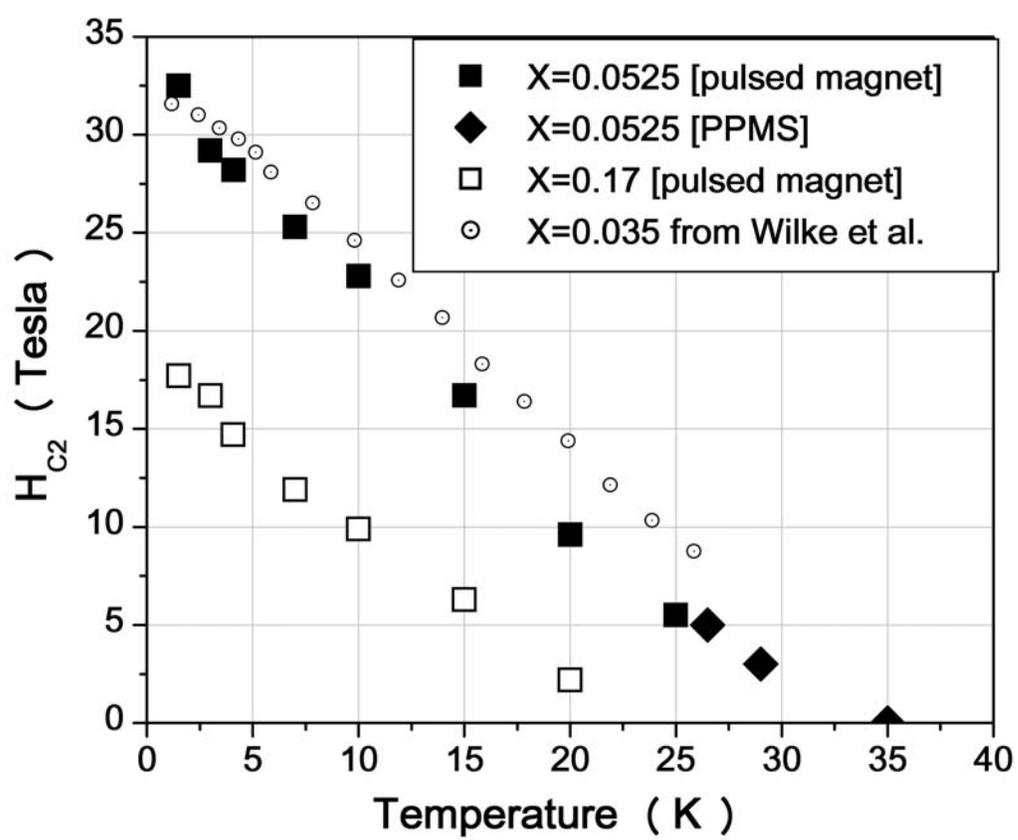

Figure 2

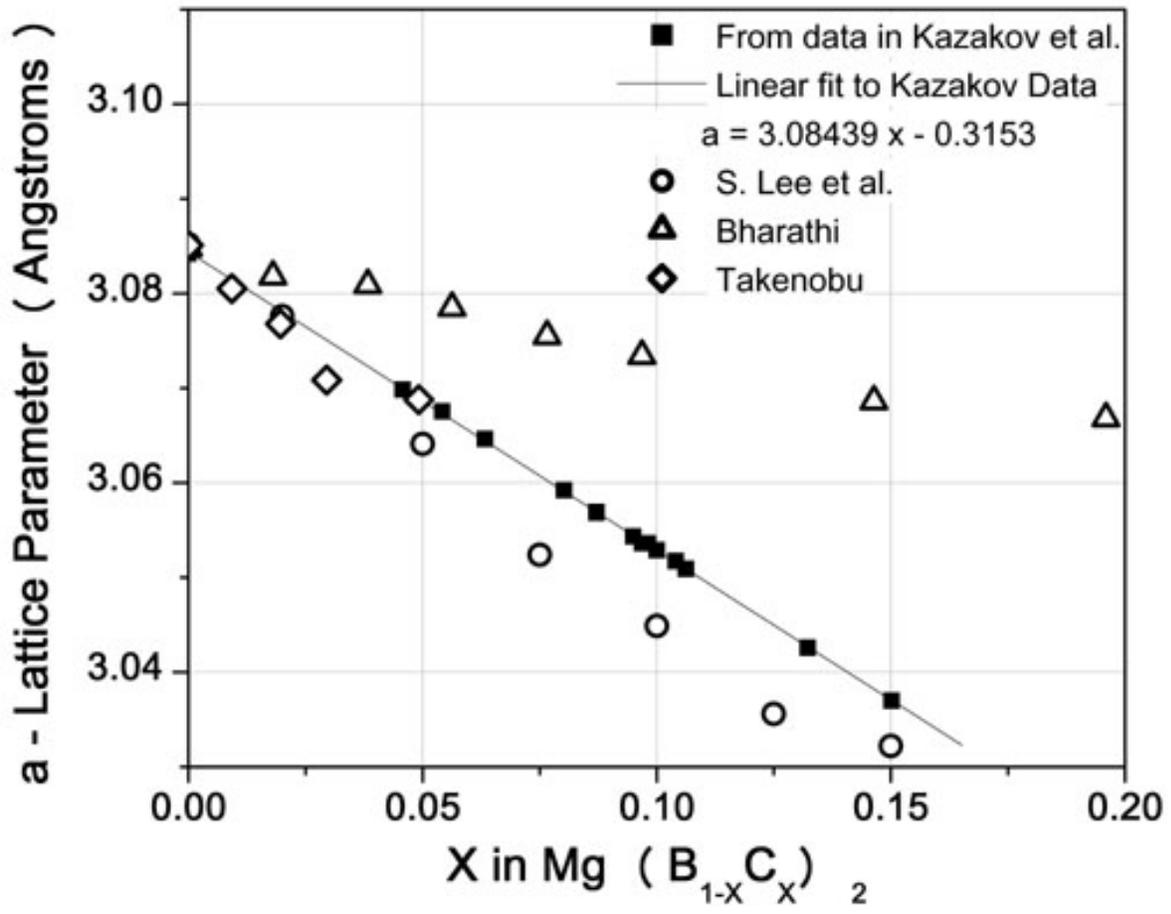

Figure 3

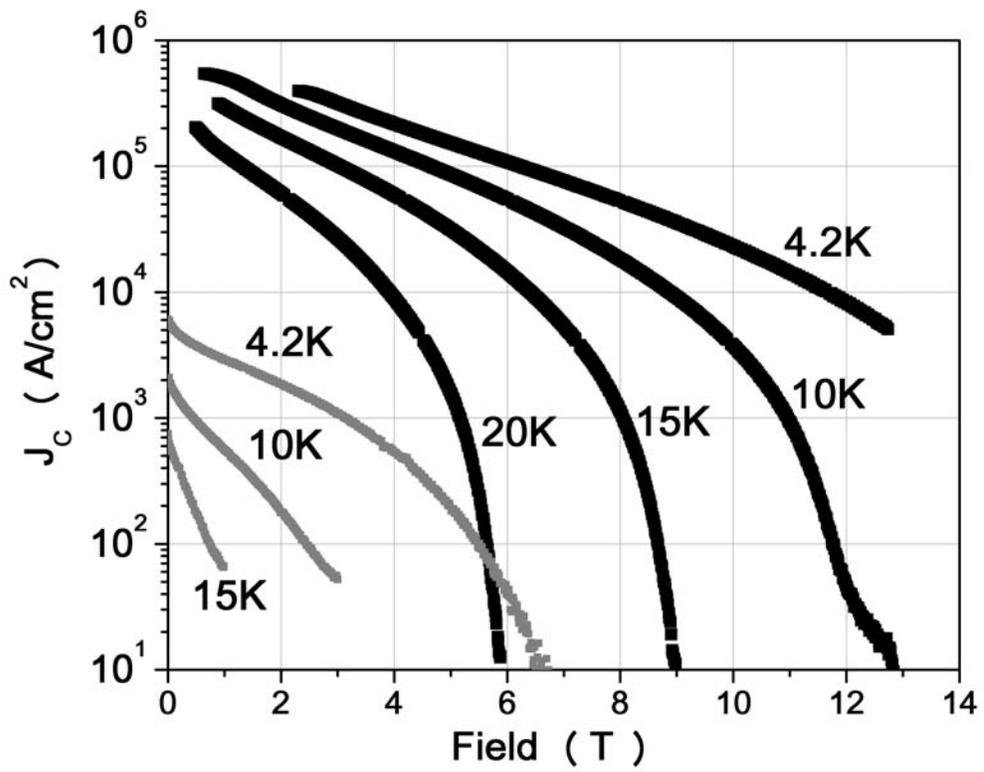